\documentclass[conference,9pt]{IEEEtran}
\usepackage{cite}
\usepackage{amsmath,amssymb,amsfonts}
\usepackage{graphicx}
\usepackage{textcomp}
\usepackage{xcolor}
\usepackage[hyphens]{url}
\usepackage[hidelinks]{hyperref}
\usepackage{booktabs}

\usepackage{tikz}
\usepackage{pgfplots}
\pgfplotsset{compat=1.18}

\usepackage{listings}
\definecolor{codegreen}{rgb}{0,0.6,0}
\definecolor{codegray}{rgb}{0.5,0.5,0.5}
\definecolor{codepurple}{rgb}{0.58,0,0.82}
\definecolor{backcolour}{rgb}{0.95,0.95,0.92}
\lstdefinestyle{codestyle}{
    backgroundcolor=\color{backcolour},   
    commentstyle=\color{codegreen},
    keywordstyle=\color{magenta},
    numberstyle=\tiny\color{codegray},
    stringstyle=\color{codepurple},
    basicstyle=\ttfamily\footnotesize,
    breakatwhitespace=false,         
    breaklines=true,                 
    captionpos=b,                    
    keepspaces=true,                 
    numbers=left,                    
    numbersep=5pt,                  
    showspaces=false,                
    showstringspaces=false,
    showtabs=false,                  
    tabsize=2,
    escapechar=\%,
    columns=fullflexible,
}
\usepackage{algorithm}
\usepackage{algpseudocode}
\usepackage{pifont}
\usepackage{enumitem}
\usepackage{multirow}

\newcommand{\mvit}{\texorpdfstring{M\textsuperscript{3}ViT}{M³ViT}}
\def\BibTeX{{\rm B\kern-.05em{\sc i\kern-.025em b}\kern-.08em
    T\kern-.1667em\lower.7ex\hbox{E}\kern-.125emX}}

\title{Edge-MoE: Memory-Efficient Multi-Task Vision Transformer Architecture with Task-level Sparsity via Mixture-of-Experts}
\author{\IEEEauthorblockN{Rishov Sarkar\textsuperscript{1}, Hanxue Liang\textsuperscript{2}, Zhiwen Fan\textsuperscript{2}, Zhangyang Wang\textsuperscript{2}, Cong Hao\textsuperscript{1}}
\IEEEauthorblockA{\textsuperscript{1}School of Electrical and Computer Engineering, Georgia Institute of Technology\\
\textsuperscript{2}School of Electrical and Computer Engineering, University of Texas at Austin\\
\href{mailto:rishov.sarkar@gatech.edu}{\nolinkurl{rishov.sarkar@gatech.edu}}, \href{mailto:lhx92505991@gmail.com}{\nolinkurl{lhx92505991@gmail.com}}, \{\href{mailto:zhiwenfan@utexas.edu}{\nolinkurl{zhiwenfan}}, \href{mailto:atlaswang@utexas.edu}{\nolinkurl{atlaswang}}\}\nolinkurl{@utexas.edu}, \href{mailto:callie.hao@gatech.edu}{\nolinkurl{callie.hao@ece.gatech.edu}}}}

\begin{document}
\maketitle
\thispagestyle{plain}
\pagestyle{plain}

\begin{abstract}

The computer vision community is embracing two promising learning paradigms: the Vision Transformer (ViT) and Multi-task Learning (MTL). 
ViT models show extraordinary performance over traditional convolution networks but are commonly recognized as computation-intensive, especially the self-attention with quadratic complexity.
MTL uses one model to infer multiple tasks %
with better performance by enforcing shared representation among tasks, but a huge drawback is that, most MTL regimes require activation of the entire model even when only one or a few tasks are needed, causing significant computing waste.
\mvit{} is the latest multi-task ViT model that introduces mixture-of-experts (MoE), where only a small portion of subnetworks (``experts'') are sparsely and dynamically activated based on the current task. \mvit{} achieves better accuracy and over 80\% computation reduction and paves the way for efficient real-time MTL using ViT.
  
Despite the algorithmic advantages of MTL, ViT, and even \mvit{}, there are still many \textit{challenges} for efficient deployment on FPGA. For instance, in general Transformer/ViT models, the self-attention is known as computational intensive and requires high bandwidth. In addition, softmax operations and the activation function GELU are extensively used, which unfortunately can consume more than half of the entire FPGA resource (LUTs). In the \mvit{} model, the promising MoE mechanism for multi-task exposes new challenges for memory access overhead and also increases resource usage because of more layer types.

To address these challenges in both general Transformer/ViT models and the state-of-the-art multi-task \mvit{} with MoE, we propose \textbf{Edge-MoE}, the \textit{first end-to-end} FPGA accelerator for \textit{multi-task ViT} with a rich collection of architectural innovations. \underline{First}, for general Transformer/ViT models, we propose (1) a novel reordering mechanism for self-attention, which reduces the bandwidth requirement from proportional to constant regardless of the target parallelism; (2) a fast single-pass softmax approximation; (3) an accurate and low-cost GELU approximation, which can significantly reduce the computation latency and resource usage; and (4) a unified and flexible computing unit that can be shared by almost all computational layers to maximally reduce resource usage.
\underline{Second}, for the advanced multi-task \mvit{} with MoE, we propose a novel patch reordering method to completely eliminate any memory access overhead.
\underline{Third}, we deliver \textit{on-board implementation and measurement} on Xilinx ZCU102 FPGA, with verified functionality and open-sourced hardware design, which achieves 2.24$\times$ and 4.90$\times$ better energy efficiency comparing with GPU (A6000) and CPU (Xeon 6226R), respectively.
A real-time video demonstration of our accelerated multi-task ViT on an autonomous driving dataset is available on GitHub,\footnote{\url{https://github.com/sharc-lab/Edge-MoE}} together with our FPGA design using High-Level Synthesis, host code, FPGA bitstream, and on-board performance results.

\end{abstract}

\section{Introduction}

The Vision Transformer (\textbf{ViT})~\cite{dosovitskiy2020image,pmlr-v139-touvron21a,liu2021swin,carion2020end} has enjoyed great popularity in the computer vision field thanks to its impressive performance. 
Comparing to convolutional neural networks (CNN) which use pixel arrays, ViT divides images into fixed-size patches and treats them as ``tokens'' in NLP. Embeddings of the tokens will be learned and fed into the transformer encoder together with a positional embedding. ViT has shown extraordinary performance in object detection and semantic segmentation tasks~\cite{liu2021swin, pmlr-v139-touvron21a}.

Meanwhile, Multi-task Learning (\textbf{MTL}) is a promising scenario where a single compact algorithm can simultaneously learn many different tasks with a much smaller model size than single-task learning (STL)~\cite{vandenhende2021multi}. In addition, MTL can learn an improved feature representation by sharing representations and utilizing regularizations between related tasks~\cite{zamir2018taskonomy,ma2018modeling}.
MTL is important and yet challenging for real-world applications, especially when the model will be deployed in an environment with limited computational capability but with real-time requirement. For instance, autonomous driving~\cite{lee2021fast} requires many tasks executed on the same platform, such as lane detection, pedestrian detection, segmentation, etc., where MTL is expected to deliver real-time performance for each task as well as swift task switch. Therefore, \textit{real-time MTL with swift task switch} is in great demand for future AI systems.

\begin{figure}[t]
    \centering
    \includegraphics[width=0.49\textwidth]{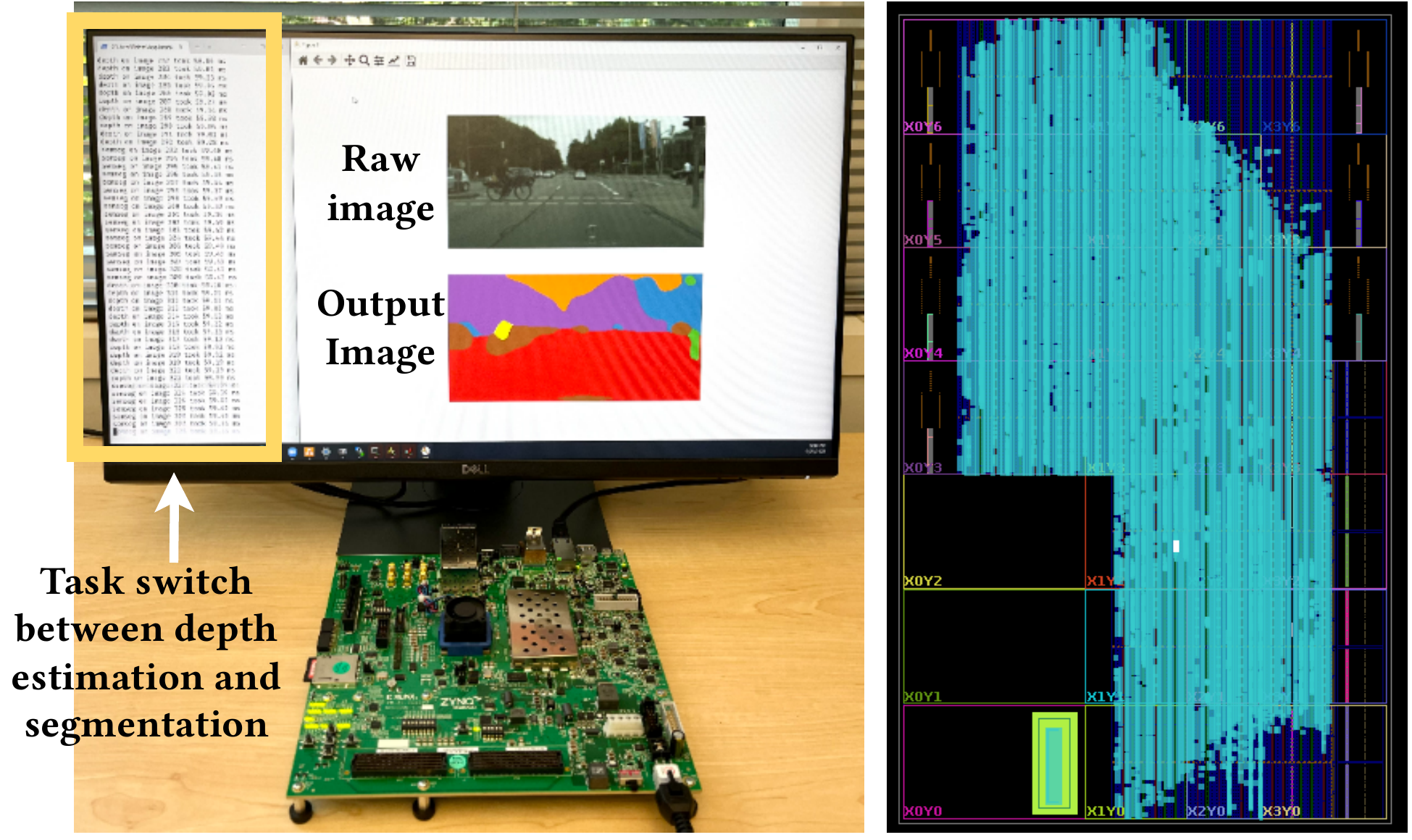}
    \caption{On-board implementation demo for our MTL ViT accelerator with guaranteed functionality and performance. The entire \mvit{} runs on the ZCU102 FPGA board; outputs are streamed to laptop only for visualization.}
    \label{fig:fpga-onboard}
\end{figure}
\begin{figure*}
    \centering
    \includegraphics[width=0.99\textwidth]{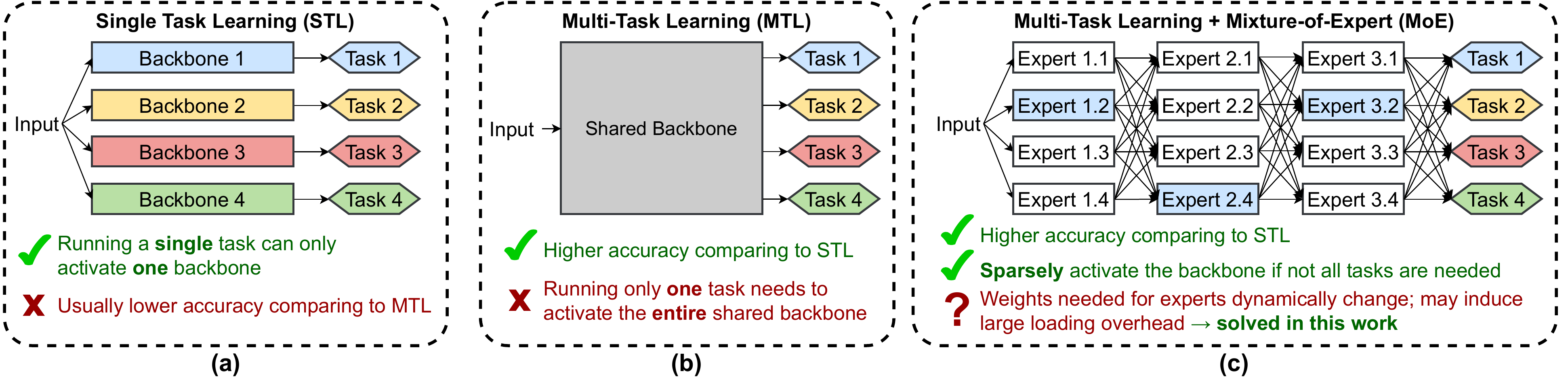}
    \caption{Pros and cons of single-task learning (STL), multi-task learning (MTL), and multi-task learning with mixture-of-expert (MoE). The most significant advantage of MTL-MoE is \textbf{sparsely activated} backbone, which saves both computation and memory footprint. The challenge of \textbf{MTL-MoE} is, the activation of ``experts'' is dynamic, depending on the current image frame. Therefore, although it largely saves the computation and memory, it may induce large weights loading overhead and cancel the benefit. In this work, we propose a novel hardware architecture for \textbf{efficient MTL-MoE with zero overhead}.}
    \label{fig:moe-motivation}
\end{figure*}

While there is a rich amount of work exploring ViT and MTL separately, applying MTL to ViT also has emerged with attractive results.
One prevailing type of MTL architectures~\cite{misra2016cross,ruder2019latent,gao2019nddr,liu2019end} adopt a shared backbone with independent task-specific head, while another type of MTL architectures~\cite{xu2018pad,zhang2019pattern,zhang2018joint,vandenhende2020mti} make task predictions with a unified decoder and then make further improvements on top of the initial predictions. These models, however, usually have a large number of FLOPs~\cite{vandenhende2021multi} and are not suitable for real-time applications on resource-constrained or latency-sensitive systems.
Chen et. al~\cite{chen2021pre} propose a pre-trained multi-task ViT for image processing, composed of multiple pairs of head and tail and a shared transformer structures across the tasks, and
Park et. al~\cite{park2022multi} propose a multi-task vision transformer for COVID-19 diagnosis and severity quantification.
A most recent work \mvit{}~\cite{liang2022m} introduces mixture-of-expert (MoE) into MTL ViT.
While existing MTL models need to activate the entire backbone unconditionally even only one or a few tasks are needed~\cite{vandenhende2021multi, misra2016cross,gao2019nddr,liu2019end}, \mvit{} can dynamically and sparsely activate only a small portion of experts, selected by a gating network, for a specific task.
\mvit{} not only achieves state-of-the-art accuracy on multi-task datasets, but also greatly reduces the model size by more than 80\% comparing to other models with similar accuracy, making it appealing for hardware deployment.
Fig.~\ref{fig:moe-motivation} illustrates the differences between single-task learning (STL), multi-task learning (MTL), and MTL+MoE~\cite{liang2022m}.

Despite the advancements in MTL with smaller model size and the sparsely activated experts in \mvit{}, there are still many \textbf{challenges} to achieve real-time inference for multi-task ViT models on FPGA.
\underline{First}, transformers, including ViT models, are notoriously known for the quadratic complexity of self-attention layer computation~\cite{dosovitskiy2020image, zhang2021algorithm, chen2022enabling}: with $N$ tokens in total, each token needs to compute attention factors with all $N$ tokens including itself. The self-attention computation has became a major bottleneck even for models with smaller size such as DeiT~\cite{touvron2021training} and T2T-ViT~\cite{yuan2021tokens}.
\underline{Second}, although \mvit{} proposes sparsely activated MoE with largely reduced computation, it introduces new challenges to memory access: since experts are dynamically determined on-the-fly, one has to either store all the weights for all experts on-chip with a large memory overhead, or load the required expert only with heavy off-chip data movement.
\underline{Third}, in Transformers and ViT models, softmax operations are extensively used after each attention layer (as opposed to convolution networks where softmax is only used in the final output), which requires a huge amount of non-linear FPGA-unfriendly computations that consume a large amount of resource and execution time (more details in Sec.~\ref{sec:challenge-softmax}). 
\underline{Fourth}, in many state-of-the-art Transformers and ViT models, a new type of activation function, GELU (Gaussian Error Linear Unit)~\cite{hendrycks2016gaussian}, has been widely used to improve training efficiency and accuracy~\cite{dosovitskiy2020image,touvron2021training,tolstikhin2021mlp,arnab2021vivit,han2021transformer}; however, its non-linearity introduces a large resource overhead and an obvious accuracy drop on FPGA (more details in Sec.~\ref{sec:challenge-gelu}).

\begin{table}[t]
    \centering
    \caption{List of our proposed techniques and their applicable models and benefits. (Transf.: Transformer; ViT: Vision Transformer; \mvit{}: multi-task ViT with mixture-of-expert; Lat.: Latency; Res: Resource)}
    \small
    \setlength{\tabcolsep}{3pt}
    \begin{tabular}{l|c|c|c|c}
    \toprule
    \textbf{Proposed Techniques}     & \textbf{Transf.} & \textbf{ViT} & \textbf{\mvit{}} & \textbf{Benefit}\\ \midrule
    \ding{202} Attention reordering     &  \ding{52} & \ding{52} & \ding{52} & Lat. \\
    \ding{203} Softmax approximation     &  \ding{52} & \ding{52} & \ding{52} & Res., Lat.\\
    \ding{204} GELU approximation     &  \ding{52} & \ding{52} & \ding{52} & Res. \\
    \ding{205} Unified computing unit & \ding{52} & \ding{52} & \ding{52} & Res., Lat. \\
    \ding{206} Patch reordering (MTL) & -- & -- & \ding{52} & Res., Lat. \\
    
    \bottomrule
    \end{tabular}
    \label{tab:list-of-techniques}
\end{table}

To address these challenges, we propose \textbf{Edge-MoE}, a hardware accelerator with innovative architectural techniques, to deliver real-time performance for multi-task ViT model \mvit{}, though the proposed techniques are generally applicable to standard Transformer/ViT models.
Table~\ref{tab:list-of-techniques} summarizes the proposed techniques, their applicable models, and the benefits.
We summarize the contributions as follows:
\begin{itemize}[leftmargin=*, itemsep=0mm]
    \item To the best of our knowledge, Edge-MoE is the \textit{first end-to-end} accelerator for \textit{multi-task} Vision Transformer, with \textit{on-FPGA implementation and measurement}, verified functionality, and open-sourced hardware design using High-Level Synthesis (HLS). Fig.~\ref{fig:fpga-onboard} depicts our on-board implementation with real-time performance on an autonomous driving dataset; a full video clip is available on GitHub.\footnote{\url{https://github.com/sharc-lab/Edge-MoE/raw/main/demo.mp4}}
    
    \item For general Transformer/ViT models with common challenges (e.g., heavy self-attention, softmax, and GELU), we propose a collection of innovative techniques, including: \ding{202} a novel attention reordering mechanism to reduce the required bandwidth from proportion to constant; \ding{203} a single-pass softmax approximation to achieve both high accuracy and fast computation speed; \ding{204} an accurate and low-cost GELU approximation with extremely low hardware resource; and \ding{205} a unified and flexible computing unit that can be shared by almost all linear layers, which drastically reduces the resource usage and thus leads to significant speedup.
    The proposed techniques can be directly applied to any Transformer/ViT model for resource and latency reduction.
    
    \item For the advanced multi-task ViT model with mixture-of-expert, \mvit{}~\cite{liang2022m}, we propose \ding{206} a novel patch reordering mechanism to eliminate memory overhead from off-chip data movement, achieving zero-overhead for expert switching and task switching. By deploying a \mvit{} on a single FPGA, we demonstrate that it is highly possible for energy-efficient and real-time multi-task ViTs to run on edge devices.
    
    \item Since \mvit{} is the state-of-the-art ViT model having all components in general Transformers and ViT models and also equipped with advanced MTL and MoE, we use it as our case study without losing any generality. Our accelerator achieves nearly 30 frames per second, 2.24$\times$ better energy efficiency than GPU (RTX A6000), and 4.90$\times$ better than CPU (Xeon 6226R). Results are measured on Xilinx ZCU102 FPGA evaluation board under 300 MHz.
    
    \item Our proposed techniques are \textit{not specific} to FPGAs but can be applied to ASIC designs as well. We use FPGAs only for concrete evaluation.
\end{itemize}

\section{Preliminary and Related Work}

\subsection{Vision Transformers and \mvit{}}

The Vision Transformer (\textbf{ViT}) is first proposed by Dosovitskiy et. al~\cite{dosovitskiy2020image} by adapting Transformers in Natural Language Processing (NLP) to processing images for computer vision tasks.
Similar to the ``tokens'' in Transformers, each image is first split into ``patches'', where each patch is of size $P\times P$ and has $P^2$ pixels.
Then patches will be flattened into vectors, projected to linear embeddings, and fed into to a standard transformer encoder in sequential order, usually with positional embeddings. Each block of the encoder is usually composed of a self-attention layer, normalization layers, a multi-layer perceptron (MLP) layer, and activation layers. In the self-attention layer (blue block in Fig.~\ref{fig:m3vit} left), 
we denote the input patch embeddings by $X\in \mathbb{R}^{L\times d}$ where $d$ is the patch embedding dimension and $L$ is the patch count. Three matrices, $Q$, $K$, and $V$, can be computed as: $Q = W^Q X$, $K = W^K X$, $V=W^V X$, where $W^Q$, $W^K$, and $W^V$ are weights. Finally, output $Y = \text{softmax}(QK^T)V \in \mathbb{R}^{L \times d}$.
This self-attention has a quadratic complexity in the softmax, where each patch needs to compute its attention score with all $n$ patches.

On top of ViT, \textbf{\mvit{}}~\cite{liang2022m} is the state-of-the-art \textit{multi-task} ViT model, which innovatively introduces a mixture-of-expert (MoE) mechanism to sparsely activate only a small portion of the model for a certain task, aiming to reduce the computation complexity.
Fig.~\ref{fig:m3vit} left illustrates the model structure of \mvit{}.
Following the general ViT architecture, \mvit{} consists of a patch embedding module followed by twelve repeated blocks.
Inside each block after the self-attention, however, there are two choices: even blocks use a traditional ViT block (yellow block in Fig.~\ref{fig:m3vit} left), while odd blocks use an MoE block (pink block). A traditional \textbf{ViT block} is composed of two fully connected layers with a GELU activation function~\cite{hendrycks2016gaussian}.
An \textbf{MoE block} is a collection of $m$ ``experts'', where each expert is a smaller MLP. The resultant output of an MoE layer is the summation of the selected top $k$ experts from $m$ expert candidates using a task-specific gating network. For instance, in the example shown in the figure, there are two gating networks for two tasks. When task A is being executed, it activates expert 1 and $m-1$, while other experts are not computed; when task B is being executed, it activates expert 1 and $m$ and others are disabled.
The advantage of this MoE approach is to \textit{sparsely} activate the experts to largely reduce computation. Specifically, in real-world multi-task model inference, not all tasks are always needed. 
With MoE, only a small portion of task-related experts are selected for computation. Without MoE, however, the entire model must always be computed.

\subsection{Existing ViT Accelerators on FPGA}

Several prior studies target the acceleration of Transformer-based models on FPGA, such as VAQF~\cite{vaqf}, SPViT~\cite{spvit}, FTRANS~\cite{ftrans}, Qi \textit{et al.}~\cite{qi2021accommodating}, Peng \textit{et al.}~\cite{peng2021accelerating}, and Auto-ViT-Acc~\cite{li2022auto}. These works note that Transformer models are computation- and memory-intensive and are too large to fit on an FPGA. Some solutions adopt various lossy model compression techniques, such as activation quantization, token pruning, block-circulant matrices (BCM) for weights, block-balanced weight pruning, and column-balanced block weight pruning. All of these require the use of compression-aware training to avoid a drop in model accuracy.
For instance, VAQF is a low-bit quantization algorithm for ViT and uses 1-bit for activation and 6-bit for weights, while Auto-ViT-Acc is a hybrid quantization framework which can automatically search for the best quantization scheme.
Tackling the ViT acceleration from different angles, our hardware novelties \textit{do not rely on model compression or re-training} and, in fact, are orthogonal to many of the compression techniques proposed in prior works, which can be applied together with our proposed techniques.

In addition, instead of focusing on the entire Transformer/ViT model, Zhang \textit{et al.}~\cite{zhang2021algorithm} propose an FPGA-based self-attention accelerator by weight pruning, while Lu et. al~\cite{lu2020hardware} propose an systolic-array based design for attention layer implementation.
By contrast, we propose an \textit{end-to-end fully functional ViT model implementation}, which exposes additional challenges that are concealed by looking at only part of the model.

For \textit{multi-task learning}, the state-of-the-art \mvit{}~\cite{liang2022m} excels in both high algorithm accuracy and low computation complexity, thanks to the MoE mechanism. It suggests a hardware-friendly method for MoE computation, but unfortunately, it only provides an optimistic estimation using FPGA without any actual hardware implementation. Therefore, our work is, to our best knowledge, the \textit{first FPGA accelerator for a multi-task ViT model with MoE}.

\section{Challenges}

\begin{figure*}
    \centering
    \includegraphics[width=\linewidth]{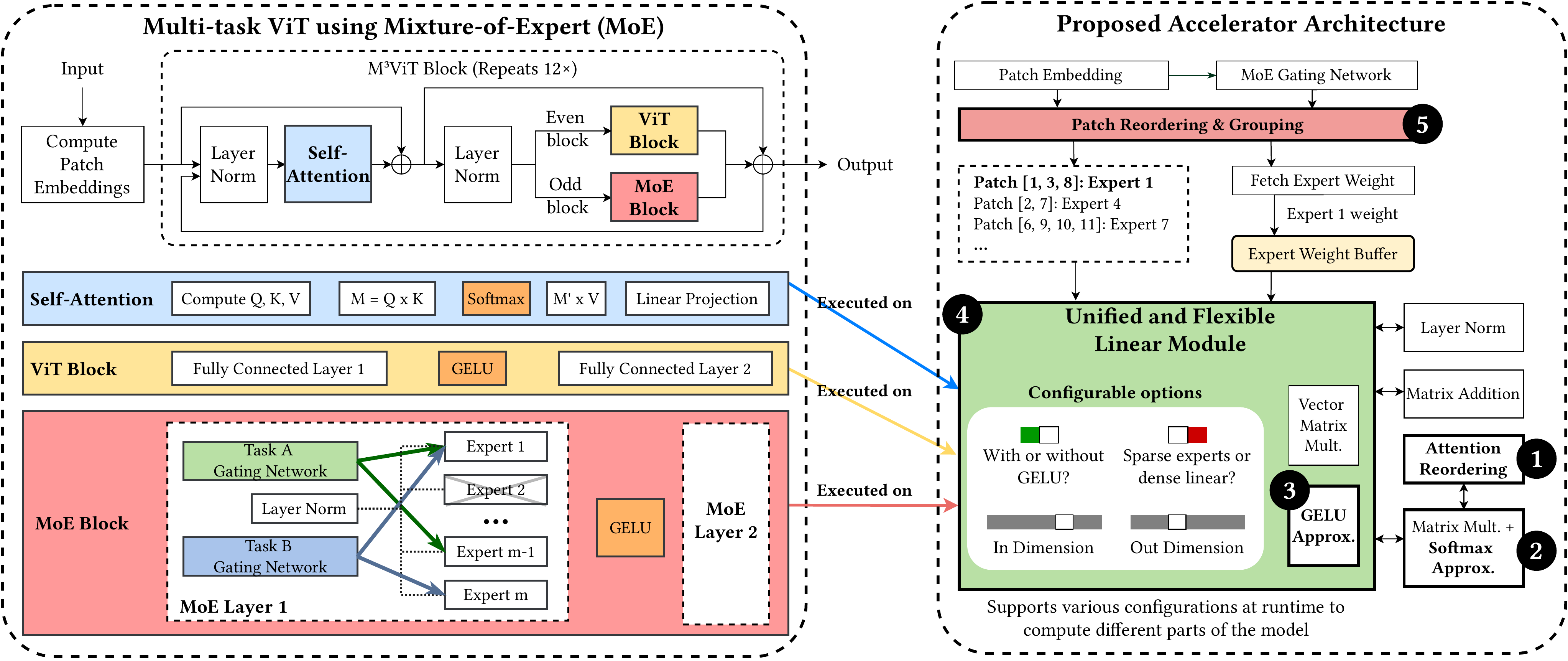}
    \caption{Left: the multi-task \mvit{} architecture with mixture-of-expert (MoE)~\cite{liang2022m}. It consists of a patch embedding module followed by 12 blocks, each of which contains a self-attention module followed by either a ViT block (on even-numbered blocks) or an MoE block (on odd-numbered blocks). Right: our proposed FPGA architecture with novel techniques, labeled by \ding{202}$\sim$\ding{206}. \ding{202}$\sim$\ding{205} is applicable to general Transformer/ViT models; \ding{206} is specialized for multi-task MoE inside \mvit{}. }
    \label{fig:m3vit}
\end{figure*}
\begin{figure}
    \centering
    \includegraphics[width=\linewidth]{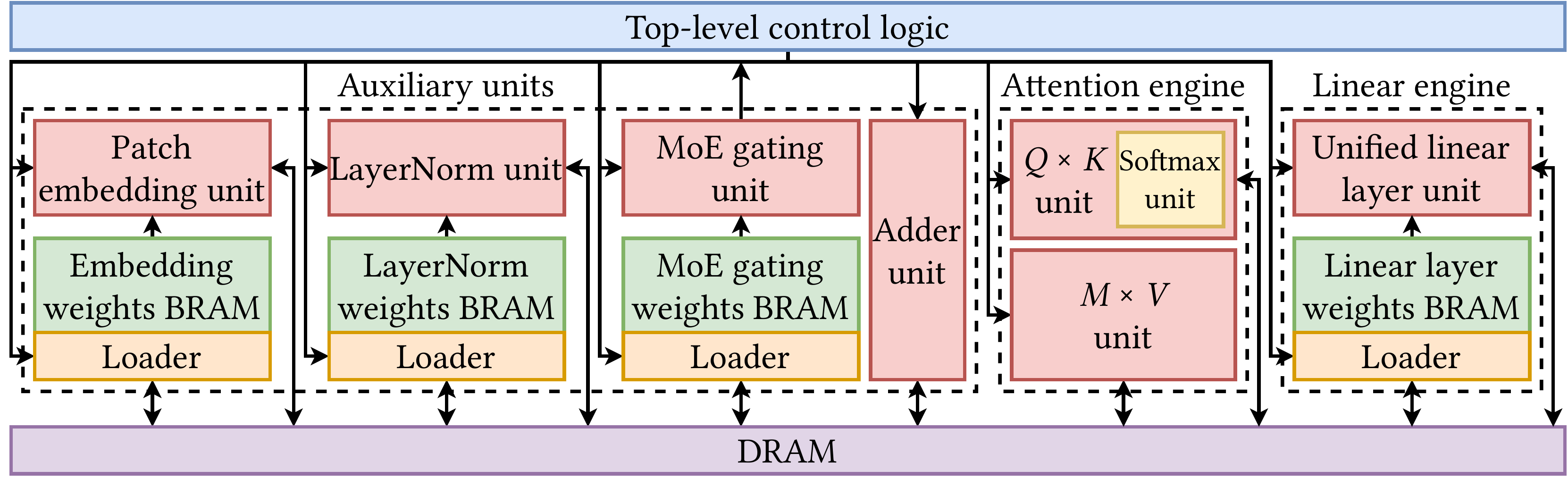}
    \caption{The system architecture of Edge-MoE.}
    \label{fig:arch}
\end{figure}

Despite the algorithmic advantages of ViT models and the computational sparsity of \mvit{} for multi-task learning, there are still severe challenges to actually deploy ViT and \mvit{} models on FPGA to achieve real-time performance, where many of them are being overlooked by existing studies. We highlight a few from two perspectives:
challenges that are \textit{general to almost all Transformer/ViT models}, and that are \textit{specific to multi-task \mvit{}}.

\subsection{General Challenges for Transformer/ViT}

\subsubsection{High Bandwidth Requirement for Self-attention}
\label{sec:challenge-attention}
It is already well-known that self-attention is computation-intensive~\cite{zhang2021algorithm, lu2020hardware} and is bandwidth constrained. Without repeating previous statements, we highlight that, achieving low latency either requires an extremely high bandwidth or large on-chip memory to buffer as many weights as possible. Neither is realistic for edge devices with limited resources.

\subsubsection{Massive Softmax Operations with Unwanted Overflow}
\label{sec:challenge-softmax}

Transformer architectures, including \mvit{}, make extensive use of the softmax operators as part of the attention mechanism to compute attention scores for each pair of tokens. In addition, on top of traditional Transformers, \mvit{} also requires softmax to compute relative expert weights from the gating network for each token in the Mixture-of-Experts block. The non-linear exponential function inside softmax is extremely hardware unfriendly (and is unfortunately heavily used in Transformer/ViT): low-precision computation may introduce a large error, while high-precision computation consumes a significant amount of resource and latency.

Furthermore, overflows in the exponential function result in catastrophic errors in softmax results. Since the softmax score for every element is dependent on the sum of \(\exp(\cdot)\) for every other element, singular overflows typically lead to massive widespread errors, which are further propagated through the next iterations of the Transformer's attention mechanism. As a result, downstream task accuracy drops to almost zero.

Since the exponential terms \(\exp(\cdot)\) must be computed many times for softmax $ \frac{\exp(x_i)}{\sum_{j=1}{N}\exp(x_j)}$, it is preferable to use fixed-point datatype on FPGA. However, since its value grows exponentially, it can easily overflow its representable range even when the argument is small and introduce a large computation error. For instance, \(\exp(7) \approx 1096.63\), which is already too large to be represented in any signed fixed-point datatype with fewer than 12 integer bits.

\subsubsection{Expensive Hardware Cost of GELU Activation}
\label{sec:challenge-gelu}

Modern Transformers and ViT models, including \mvit{}, make use of the GELU (Gaussian Error Linear Unit) activation function~\cite{hendrycks2016gaussian} for fully connected layers and MLPs instead of ReLUs.
GELU can be regarded as a smoother ReLU but has more expressiveness because of its better non-linearity and leads to faster and better convergence of neural networks~\cite{devlin2018bert, liu2021swin, liang2022m, dosovitskiy2020image}.
GELU$(x)$ is computed as follows:
\begin{equation}
    \text{GELU}(x) = x\Phi(x) = x\cdot 0.5(1 + \text{erf}(x / \sqrt{2}))
    \label{eq:gelu}
\end{equation}
where $\Phi(x)$ is the standard Gaussian cumulative distribution function and \(\mathrm{erf}(\cdot)\) is the Gauss error function.

Implementing GELU activation using Eq.~\eqref{eq:gelu} is extremely expensive on FPGA: it requires 161.8k LUTs (59\% of those available on the ZCU102) for just a single instance using 32-bit fixed-point datatype.
Therefore, one must use approximation for GELU.

Given the similarity of \(\mathrm{erf}(\cdot)\) and \(\tanh(\cdot)\), it is suggested that GELU can be approximated using \(\tanh(\cdot)\) as follows~\cite{hendrycks2016gaussian}:
\begin{equation}
    \text{GELU}(x) \approx 0.5x(1+\text{tanh}(\sqrt{2/\pi}(x+0.044715x^3)))
    \label{eq:gelu-tanh}
\end{equation}

However, this still consumes significant hardware resources: only one instance consumes 18.7k LUTs (6\% of available LUTs on ZCU102), while there are thousands of them expected to be computed in parallel. The same work also suggests an approximation using the sigmoid function, which takes much fewer resources, 4.7k LUTs (1\%) for one instance, but is significantly less accurate.

Therefore, it is challenging to maintain both high approximation accuracy and low computation complexity and hardware resource in computation of the GELU activation function.

\subsubsection{Duplication of Linear Layer Logic}
\label{sec:challenge-linear}
Transformers and ViT models extensively use fully connected linear layers, including in MLP, patch embedding for self-attention, and the linear projection at the end of self-attention.
Creating a dedicated hardware module for each linear operation consumes too many resources and limits parallelism, and thus aggressive resource sharing across the linear layers in different module blocks is needed.
However, though the computational nature of the linear layers is the same, their structures are different when appear in different blocks. Variations include: input and output dimensions, weight fetch and store addresses, preferred data parallelism mechanism (partition), and whether to use activation functions. %

The \mvit{} model aggravates the challenge by introducing another block type: the MoE block in addition to the traditional ViT block (see Fig.~\ref{fig:m3vit} left).
The ViT blocks use MLPs with a larger hidden dimension than that of the MLPs used in the MoE blocks. Furthermore, due to the nature of mixture-of-experts computation, expert MLPs do not process all tokens; instead, they must accept sparse inputs of only a subset of input tokens.

Therefore, designing a unified computing unit to accommodate all types of linear layers to encourage resource sharing and thus to enable larger parallelism is a critical challenge.

\subsection{Model Specific Challenges for MTL \mvit{}}

\subsubsection{Memory Accesses for Mixture-of-Experts}
\label{sec:challenge-moe}

A defining feature of the \mvit{} algorithm is its use of mixture-of-experts (MoE) to sparsify a large ViT model by selecting, for each input token, a different set of ``experts'' to use to compute its output representation. A gating network is used to score each of \(m\) experts with respect to each token, and the experts with the highest \(k\) scores are used for that token's output.
Fig.~\ref{fig:expert-reorder}(a--b) depicts how the input image is split into patches and how each patch selects a subset of experts for its computation.

Following the computation of the gating network, a natural approach to perform the subsequent MoE computation is to treat it similarly to any other MLP: first, load all weights for all experts into on-chip BRAM. Then, computation of each token's outputs is straightforward by providing the weights of each of the \(k\) selected experts from BRAM as the weight inputs to an MLP. However, this approach requires loading all \(m\) experts on-chip. In \mvit{}, $m = 16$, and whole expert weights cannot fit into the available BRAM resources.

An alternative approach is to compute the patches one by one and fetch each expert only when needed, as shown in Fig.~\ref{fig:expert-reorder}(c). 
For instance, when computing patch 1, since its selected experts are 1 and 4, expert 1's weights are first loaded, followed by expert 4; next, when computing patch 2, expert 1's weights have to be reloaded since they were swapped out.
This sidesteps the issue of limited BRAM, but it incurs severe memory delays, as the expert weights have to be reloaded constantly.

\section{Proposed Methods}

In this section, we describe our novel solutions to alleviate the previously mentioned challenges.
The proposed overall architecture is presented in Fig.~\ref{fig:m3vit} with our proposed novel techniques, labeled from \ding{202} to \ding{206}, where \ding{202}$\sim$\ding{205} are generally applicable to standard Transformers and ViT models, and \ding{206} is specific to the multi-task \mvit{} with MoE. Fig.~\ref{fig:arch} shows another perspective, including memory hierarchy.

The proposed design is a \textit{fully functional end-to-end} accelerator including the following major components and key techniques: 
\begin{itemize}[leftmargin=*, itemsep=0mm]
    \item Initial patch embedding computation;
    \item A task-specific gating network to generate expert selection;
    \item Patch reordering and grouping to consolidate expert weight loading and to eliminate memory access overhead (\ding{206} in Sec.~\ref{sec:expert-reorder});
    \item A unified and flexible linear module which consolidates a large portion of linear layers across the entire model (\ding{205} in Sec.~\ref{sec:unified-linear});
    \item An efficient attention reordering module for self-attention computation with largely reduced memory access and increased parallelism, overcoming the bandwidth bottleneck (\ding{202} in Sec.~\ref{sec:attention-reorder});
    \item A novel single-pass, accurate, and low-cost softmax approximation module used for self-attention (\ding{203} in Sec.~\ref{sec:softmax});
    \item An accurate and low-cost GELU approximation module being integrated into the unified linear module (\ding{204} in Sec.~\ref{sec:gelu});
    
\end{itemize}

Note that although the depicted accelerator is for \mvit{}, it is easily applicable to standard Transformers and ViT models simply by removing the MoE layer and patch reordering module.
In the experiments, we will also evaluate its performance on standard ViT models. In the following sections, we discuss our proposed key techniques.

\subsection{Attention Reorder for Bandwidth Reduction}
\label{sec:attention-reorder}

\begin{figure}
    \centering
    \includegraphics[width=\linewidth]{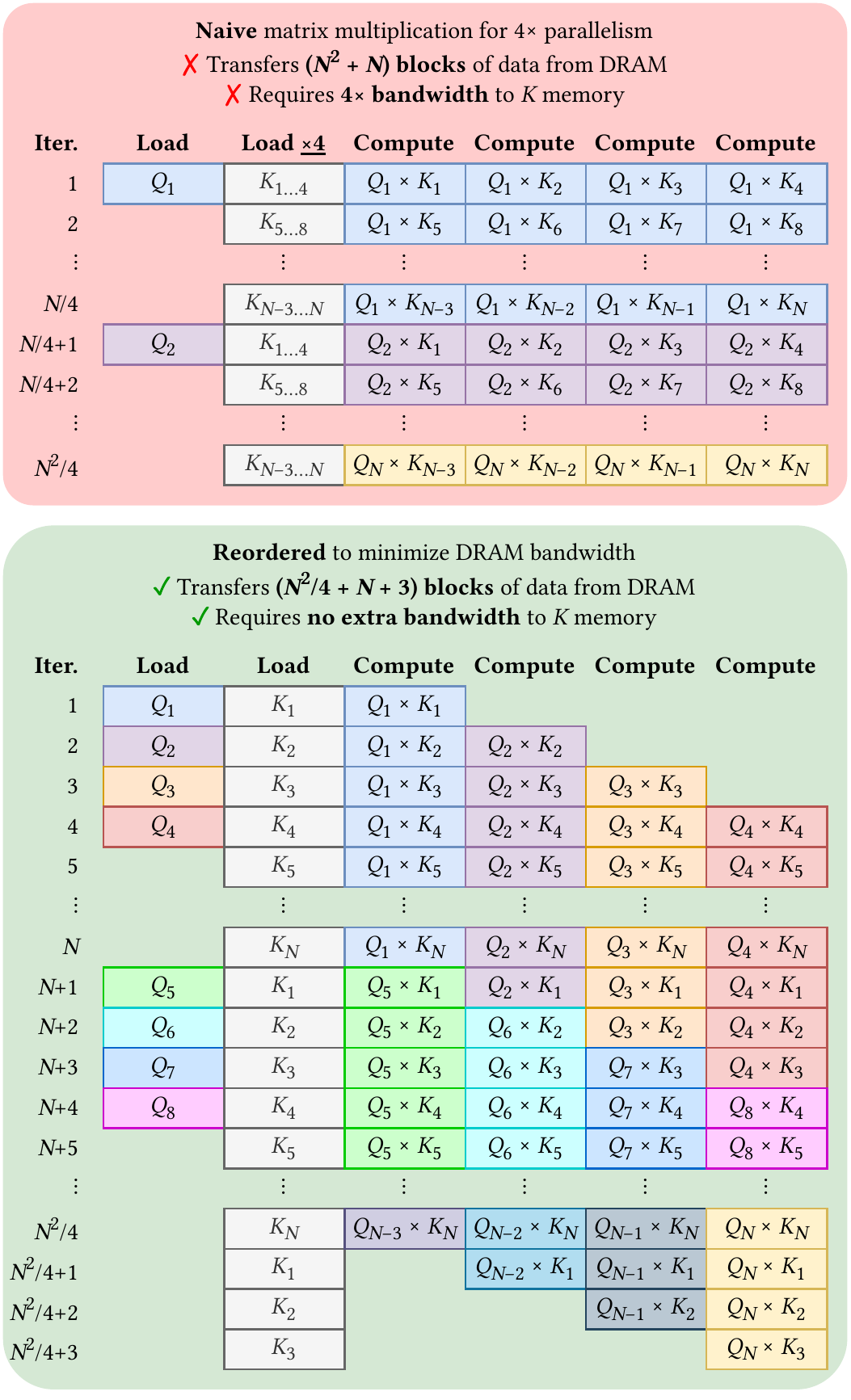}
    \caption{The proposed computation reordering for parallelism in the self-attention mechanism.}
    \label{fig:attn}
\end{figure}

\noindent
\textbf{Without reordering}.
The attention mechanism in Transformer models as well as \mvit{} is computation-intensive and bandwidth-constrained. Figure~\ref{fig:attn} (top) illustrates the bandwidth limitation by showing the computation flow of $Q_i \times K_j $ for all the tokens, where $1 \leq i \leq N$, $1 \leq j \leq N$, and $N$ is the total number of tokens.
To finish such computation for all tokens, $N^2$ pairs of vector multiplication is needed.
A straightforward way is to load $Q_i$ tokens in order, from $Q_1$ to $Q_N$; then, for each $Q_i$, we load $K_j$ tokens from $K_1$ to $K_N$, as demonstrated in Fig.~\ref{fig:attn} (top).
The limitation of this method is, increasing attention parallelism also requires an increase in DRAM bandwidth by the same factor. In this example, if one wants to achieve a parallelism of 4, then each $Q_i$ token must be multiplied by four $K_j$ tokens in every iteration; therefore, it requires the four $K_j$ tokens being loaded from DRAM every iteration, requiring four times the bandwidth to access \(K\). As a result, each $K_j$ token is loaded \(N\) times, resulting in the transfer of \(N^2\) blocks of data, and the $Q_i$ tokens are loaded once, requiring an additional \(N\) blocks.
Therefore, to compute the attention in this fashion will need $N^2+N$ times of data transfer to achieve a total latency of $N^2/p$; therefore, the \textbf{bandwidth requirement is proportional to parallelism}, approximately $p$. The memory requirement is one buffer for $Q$ token and $p$ buffers for $K$ token, totaling $p+1$ buffers.

\begin{table}
    \centering
    \caption{Data loading amount, latency, memory requirement, and bandwidth requirement with and without our proposed attention reordering under parallelism $p$.}
    \label{tab:attention-reorder}
    \small
    \setlength{\tabcolsep}{3pt}
    \begin{tabular}{c|c|c|c|c}
    \toprule
    \textbf{Approach} & \textbf{Data Load} & \textbf{Latency} & \textbf{Bandwidth} & \textbf{Memory}\\ \midrule
w/o reorder   &  $N^2 + N$ & $\tfrac{N^2}{p}$ & $\sim p$  & $p+1$ \\
w/ reorder  &  $\tfrac{N^2}{p} + N + p-1$ & $\tfrac{N^2}{p} + p-1$ & $\sim 1$ & $p+1$ \\

\bottomrule
    \end{tabular}
\end{table}

\noindent
\textbf{Our proposed reordering}.
Addressing this challenge, we introduce a novel reordering strategy, shown in Figure~\ref{fig:attn} (bottom), that \textbf{achieves arbitrary parallelism with a constant bandwidth} for the input matrices. Specifically, for a desired parallelism factor \(p\) (4 in the figure), we first 
cache one batch of \(p\) tokens of \(Q_i\) in a local buffer (e.g., $Q_1$ to $Q_4$ in the first batch); then, during each iteration, we load a new $K_j$ token and multiply it with the $p$ tokens of $Q$ in the local buffer. The memory required is $p$ buffers for $Q$ token and 1 buffer for $K$ token, also $p+1$ buffers in total.

Notice that some certain tokens of \(Q\) are not aligned with the start of the \(K\) matrix, such as \(Q_2\) in the figure. Thus certain outputs are initially ``missing,'' like \(Q_2 \times K_1\), since \(Q_2\) was not in the buffer during the iteration in which \(K_1\) was fetched from DRAM. These ``missing'' outputs are revisited after the rest of the \(K\) matrix is processed, while the next batch of \(Q\) is being read. Conveniently, the last output for each token is computed just before a new token is to be read into the buffer, thus incurring no idle time between batches.
At the end of the process, once all of \(Q\) has been read, up to \(p - 1\) iterations are added to process any remaining ``missing'' outputs in the last batch.

Since \(Q\) is processed in batches of \(p\) tokens at a time, each token of \(K\) is reused across \(p\) tokens, meaning that the whole \(K\) matrix is read only \(\frac{N}{p}\) times. This results in the transfer of \(\frac{N^2}{p}\) blocks of the \(K\) matrix, plus up to \(p - 1\) blocks additionally transferred at the end to account for missing outputs. The \(Q\) matrix is still read only once, resulting in a total transfer of \(\frac{N^2}{p} + N + p - 1\) blocks of data.
Given a total latency of $\frac{N^2}{p}+p-1$, the \textbf{required bandwidth is approximately 1 regardless the parallelism $p$}.
This reuse of the \(K\) matrix for \(p\) blocks of \(Q\) is also exactly the same reason why the baseline strategy needs \(p\) times the bandwidth for \(K\) to achieve the same parallelism.

Table~\ref{tab:attention-reorder} summarizes performance without and with our proposed reorder with parallelism $p$, including the total number of data loading, latency, required bandwidth, and on-chip memory. Apparently, without reordering, the bandwidth requirement is proportional to parallelism, while our proposed reordering mechanism reduces the bandwidth to a constant.

While we only depict the optimization on \(Q \times K\), it can be applied similarly to \(M' \times V\) \textbf{in reverse:} instead of \textit{producing} up to four elements of \(Q \times K\) in each iteration, as shown in Fig.~\ref{fig:attn} (bottom), the four elements at those indices are \textit{loaded} every iteration. They are then processed using softmax, multiplied by the \(V\) matrix (which is loaded the same way as \(K\) in Fig.~\ref{fig:attn} bottom), and accumulated into a cache of four tokens. After every \(N\) iterations, rather than \textit{loading} four tokens of \(Q\), the contents of the four-token cache are \textit{written back} to DRAM as the final result of four of the tokens in the \(M' \times V\) computation.

\begin{figure}
    \centering
    \begin{tikzpicture}
        \begin{axis}[width=\linewidth, height=4cm, xmin=-5, xmax=5, ymin=-22.5, ymax=12.5, xlabel={Bias \(b\)}, ylabel={Input \(x\)}]
        	\addplot[mark=none,color=red!25,fill] {-22.5} \closedcycle;
        	\addplot[mark=none,color=red!25,fill] {12.5} \closedcycle;
        	\addplot[mark=none,color=green!25,fill] {ln(2 ^ -23) + x} \closedcycle;
        	\addplot[mark=none,color=green!25,fill] {ln(512 - (2 ^ -22)) + x} \closedcycle;
        	\node[color=red,anchor=north west,inner sep=0pt,outer sep=0pt] at (axis cs:-4.75, 10) {\bfseries Overflows};
        	\node[color=red,anchor=south east,inner sep=0pt,outer sep=0pt] at (axis cs:4.75, -20) {\bfseries Rounds to zero};
        	\addplot[<->] coordinates {(-3, 3.23832462) (-3, -18.9423852)};
        	\node[text width=4cm, anchor=west] at (axis cs:-3, -5.85203026) {Usable range is constrained by precision and range of datatype, regardless of bias \(b\)};
        \end{axis}
    \end{tikzpicture}
    \caption{The usable range (marked in green) of the function \(\exp(x - b)\) for a 32-bit signed fixed-point datatype with 10 integer bits for different values of the bias \(b\). Too large $b$ will result in rounding to zeros, while too small $b$ will result in overflow; optimal $b$ value depends on the input $x$.}
    \label{fig:exp-bias-tradeoff}
\end{figure}
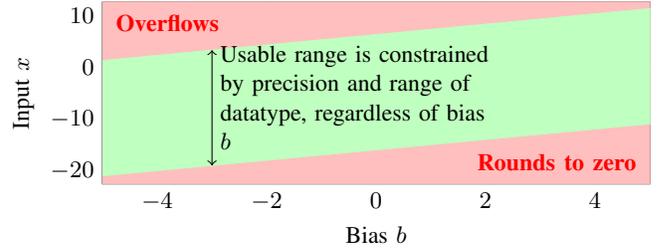

\subsection{Single-Pass Softmax Accumulation}
\label{sec:softmax}

\subsubsection{Dynamic bias for accuracy compensation}
\label{sec:dynamic-bias}

As discussed in Sec.~\ref{sec:challenge-softmax}, the biggest challenge of softmax computation on hardware is the overflow of the exponential term $\text{exp}(x)$, which largely affects the model accuracy since softmax is extensively used in Transformer and ViT models.

To address the overflow problem to compensate accuracy loss, we propose a \textit{compensation bias}, denoted by \(b\), to re-adjust the output range for $\text{exp}(x_i)$ and $\text{exp}(x_j)$ while computing $ \frac{\exp(x_i)}{\sum_{j=1}^{N}\exp(x_j)}$.
Specifically, we subtract $b$ from the arguments of both exponential functions \(\exp(x_i)\) and \(\exp(x_j)\) to reduce the magnitude of the exponential terms; thanks to the additive property of exponential functions, the compensation does not change the softmax result:
\begin{equation}
\textstyle
\frac{\exp(x_i - b)}{\sum_{j = 1}^{N}{\exp(x_j - b)}} = \frac{\exp(-b) \exp(x_i)}{\exp(-b) \sum_{j = 1}^{N}{\exp(x_j)}} = \frac{\exp(x_i)}{\sum_{j = 1}^{N}{\exp(x_j)}}    
\label{eq:softmax-with-bias}
\end{equation}

The challenge is, however, selecting larger bias value \(b\) can make the exponential term \(\exp(x - b)\) less likely to overflow, but it also introduces accuracy loss. If \(x - b\) decreases too much, the value \(\exp(x - b)\) may be too small to be accurately represented by fixed-point numbers.
Figure~\ref{fig:exp-bias-tradeoff} demonstrates this trade-off:
when the input $x$ is small, larger $b$ will make the value of \(\exp(x - b)\) round to zero;
when $x$ is large, smaller $b$ will result in overflow.
Therefore, an optimal bias $b$ heavily depends on the input value $x$, and thus \textit{a ``one-size-fits-all'' approach using a predetermined bias \(b\) is not feasible for fixed-point datatype} for softmax computation.

Therefore, we propose to \textit{dynamically} decide the bias value $b$ for each token (with embedding $x_i$), where \(b = \max_{j \in \{1, \ldots, N\}}(x_j)\) which ensures that \(\exp(x_j - b) \leq 1\) for all \(j \in \{1, \ldots, N\}\), preventing overflows while also preserving accuracy. With this dynamic bias, even if a particular term \(\exp(x_j - b)\) experiences loss of precision or rounds to zero, accuracy is still maintained. The reason is, even if \(\exp(x_j - b)\) rounds to zero, it implies that \(x_j\) is much smaller than \(\max(x_j)\), and thus its contribution to the overall softmax calculation is negligible and can be ignored without loss of accuracy.

\subsubsection{From three-pass softmax to single-pass softmax}
\label{sec:single-pass-softmax}

Using a dynamic bias, however, implies \textbf{an expensive three-pass approach} to compute Eq.~\eqref{eq:softmax-with-bias}: \textbf{Pass 1} over all tokens \(j \in \{1, \ldots, N\}\) is required to scan all the input values find the optimal bias \(b\); \textbf{Pass 2} is required to compute the sum of exponential terms in the denominator \(s = \sum_{j=1}^{N} \exp(x_j - b)\); \textbf{Pass 3} is required to compute \(\exp(x_i - b)/s\) for each token to finish the softmax computation.

\begin{algorithm}[t]
\small
    \caption{Online algorithm to compute softmax bias \(b\) and denominator summation \(s\) simultaneously}
    \begin{algorithmic}[1]
        \Require \(x\), a list of \(N\) elements
        \Ensure \(b = \max(x)\), \(s = \sum_{j = 1}^{N} \exp(x_j - b)\)
        \State \(b \gets -\infty\), \(s \gets 0\)
        \For{\(j \in \{1, \ldots, N\}\)}
            \If{\(x_j > b\)}
                \State \(s \gets s \cdot \exp(b - x_j) + 1 \) 
                \State \(b \gets x_j\)
            \Else
                \State \(s \gets s + \exp(x_j - b)\)
            \EndIf
        \EndFor
    \end{algorithmic}
    \label{alg:online-softmax}
\end{algorithm}
\begin{figure}
    \centering
    \includegraphics[width=0.4\textwidth]{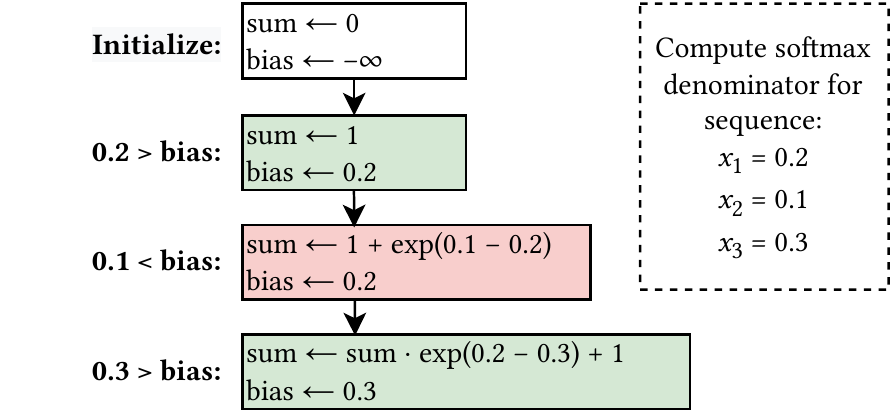}
    \caption{An example with three elements demonstrates how our online softmax algorithm works no matter the order in which the elements are sorted.}
    \label{fig:online-softmax}
\end{figure}

To reduce the computation from three-pass to one-pass to reduce the computation latency, we propose an online algorithm to compute the bias \(b\) and softmax denominator summation \(s\) simultaneously, effectively combining \textbf{Pass 1} and \textbf{Pass 2}.
The proposed process is shown in Algorithm~\ref{alg:online-softmax}. We begin by initializing the sum to 0 and the bias to the most negative value representable by the fixed-point datatype (symbolized by \(-\infty\)) (line 1). For each element \(x_j\) in the sequence, we first determine whether this element is the new maximum and thus should be the new bias. If so, we scale the current sum by \(\exp(b - x_j)\), thereby effectively updating the bias from \(b\) to \(x_j\) for all previous elements in the summation without explicitly doing so, and add 1, which is simply \(\exp(x_j - b)\) for \(b = x_j\) (lines 4--5).
Otherwise, we add \(\exp(x_j - b)\) to the sum, where \(b\) is the existing bias (line 7).
Figure~\ref{fig:online-softmax} shows an example of this algorithm on three elements, \{0.2, 0.1, 0.3\}. It shows that, while we are processing the elements one by one, both the sum and the bias $b$ are updated simultaneously in a single pass.

Finally, to avoid a separate \textbf{Pass 3} to compute the final softmax results \(\exp(x_i - b)/s\) for all the tokens, we keep the original scores \(x_i\) stored alongside the bias \(b\) and denominator \(s\). The next process that needs the softmax result, such as the self-attention subprocess that reads attention scores and multiplies them with the \(V\) matrix, includes the exponentiation and division hardware to compute the softmax result \(\exp(x_i - b)/s\) as it reads each incoming element \(x_i\). With pipelining, the added latency for the exponentiation and divison is negligible.

\subsection{Accurate Low-Cost GELU Approximation}
\label{sec:gelu}

\begin{figure}
    \centering
    \includegraphics[width=0.42\textwidth]{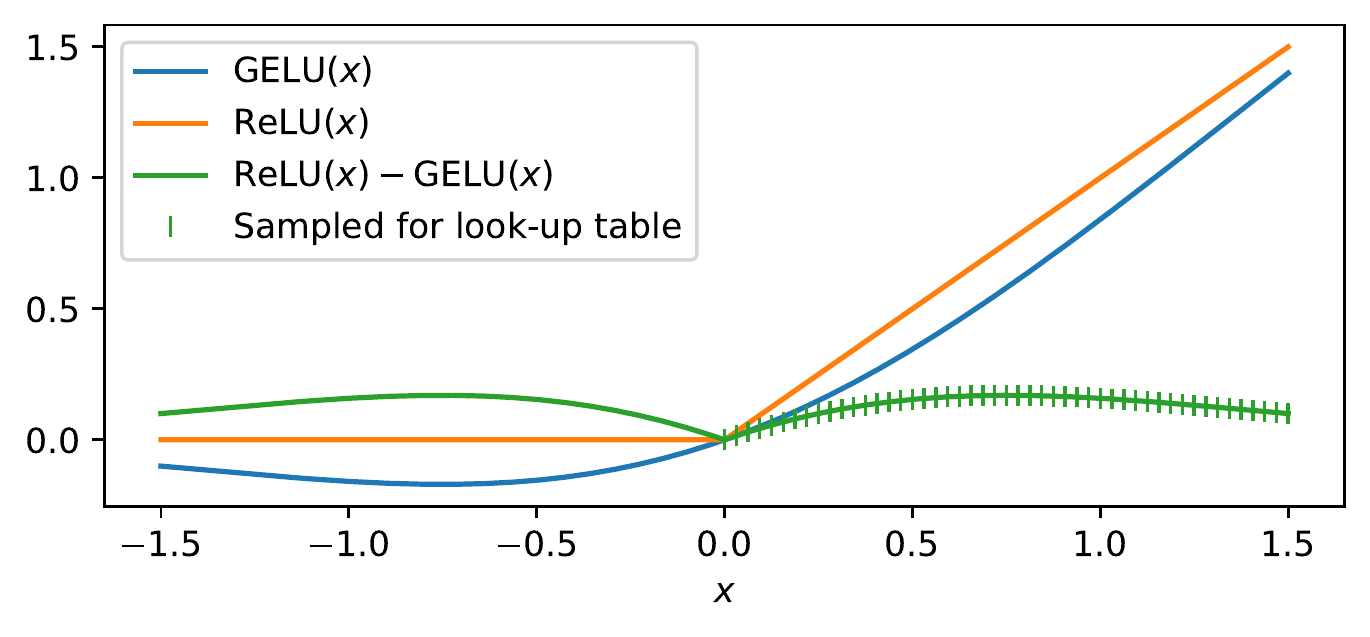}
    \caption{GELU approximation using ReLU and a calibration function $\delta(x)$. $\delta(x)$ is an even function, so we uniformly sample $\delta(x)>0$ and store the discretized values in a look-up table.}
    \label{fig:gelu-approximation}
\end{figure}

As discussed in Sec.~\ref{sec:challenge-gelu}, direct GELU computation is extremely expensive, while the existing approximation methods using $\tanh(\cdot)$ in Eq.~\eqref{eq:gelu-tanh} or sigmoid are either resource-heavy or largely inaccurate.
To address this challenge, we propose an accurate and hardware-friendly approximation for GELU with extremely low hardware cost.
The proposed approximation method includes four steps of optimization, depicted in Fig.~\ref{fig:gelu-approximation}.

\underline{First}, we notice that the value of GELU is similar to ReLU, which is easily computed as the function \(\mathrm{ReLU}(x) = \max(x, 0)\).
Therefore, we use the ReLU function as the base approximation with a small calibration value $\delta(x)$ based on the input $x$:
\begin{equation}
    \text{GELU}(x) \approx \text{ReLU}(x) - \delta(x)
\end{equation}
where $\delta(x)$ can be uniformly sampled to fixed point representations, pre-computed, and stored in look-up tables in ROM. 
Fig.~\ref{fig:gelu-approximation} shows three curves: the ReLU curve in orange, the GeLU curve in blue, and their difference $\delta$ in green.

\underline{Second}, we notice that the $\text{erf}(\cdot)$ in GELU is an odd function holding $\text{erf}(-z)=-\text{erf}(z)$. For $x>0$, we have:
\begin{gather}
\delta(x) = \text{ReLU}(x) - \text{GELU}(x) = x - x\cdot \tfrac{1}{2}(1+\text{erf}(\tfrac{x}{\sqrt{2}})) \\
\begin{split}
\delta(-x) &= 0 - (-x)\cdot \tfrac{1}{2}(1+\text{erf}(\tfrac{-x}{\sqrt{2}})) \\
&= x\cdot \tfrac{1}{2}(1-\text{erf}(\tfrac{x}{\sqrt{2}})) = \delta(x)
\end{split}
\end{gather}
Therefore, the difference between GELU and ReLU is symmetric about zero, i.e., $\delta(x)$ is an even function. This allows us store only values of \(\mathrm{ReLU}(x) - \mathrm{GELU}(x)\) where \(x \geq 0\). Fig.~\ref{fig:gelu-approximation} shows that only half of the $\delta(x)$ curve is sampled and stored on-chip.  %

\underline{Third}, to store the look-up table for discretized values for $\delta(x)$, since \(0 \leq \mathrm{ReLU}(x) - \mathrm{GELU}(x) < 1\) for all \(x \in \mathbb{R}\), we can store only the unsigned 22 fractional bits of the high-precision 32-bit fixed-point datatype used for GELU approximation, which maintains full precision without requiring storage of any of the integer bits.

\underline{Fourth}, we truncate the look-up table at the point where \(\mathrm{GELU}(x)\) rounds to \(\mathrm{ReLU}(x)\) based on the 32-bit fixed-point datatype; for any query value \(x\) outside this range, we simply use \(\mathrm{ReLU}(x)\) as a sufficiently precise approximation of \(\mathrm{GELU}(x)\). In addition, the look-up table step size is chosen to be a negative power of two. This makes the division operation required to compute the look-up table index equivalent to a bit shift, making it very efficient and cheap to implement in hardware.

\subsection{Expert-by-Expert Computation Reordering}
\label{sec:expert-reorder}

\begin{figure*}
    \centering
    \includegraphics[width=0.97\linewidth]{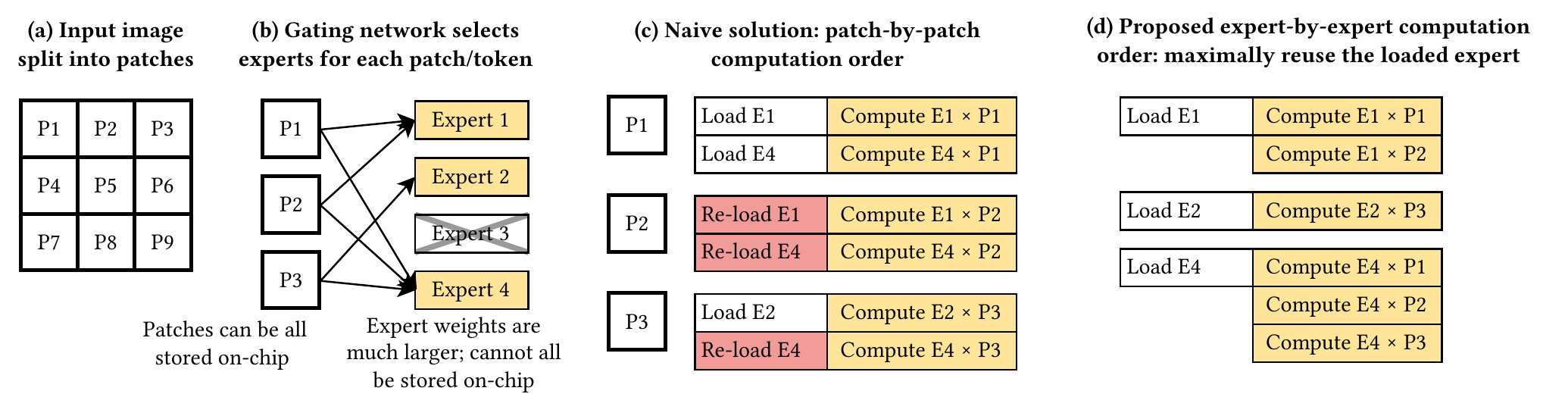}
    \caption{Our expert-by-expert computation reordering strategy (d). Comparing with the naive patch-by-patch method, the proposed method maximally reuse each loaded expert and thus each expert only needs to be loaded once. This method eliminates all potential memory and data movement overhead introduced by MoE.}
    \label{fig:expert-reorder}
\end{figure*}
As described in Sec.~\ref{sec:challenge-moe}, MoE models such as \mvit{} have an additional challenge beyond standard Transformer-based models, which is to process multiple experts in an unpredictable access pattern without constantly reloading expert weights. Our proposed method is to reorder the token-by-token computations to be performed expert-by-expert instead, as shown in Figure~\ref{fig:expert-reorder}(d). During the processing of the gating network, when the top-\(k\) expert MLPs are selected for each token, the token indices are added to per-expert queues for later processing.

We construct a metaqueue of all experts whose queues have nonzero length. This allows us to skip the loading step of any experts not used in the current MoE block, such as expert 3 in the figure.
Then, for each expert in the metaqueue, we load the expert weights and biases into our unified linear layer module and generate all outputs for that expert. The gating network assigns scores for each token/expert pairing, and each expert's output for a given token is weighted by the score before being accumulated onto the existing partial output for that token. Tokens not in the expert's previously generated queue are skipped entirely.

By pre-aggregating a queue of all tokens that an expert MLP needs to compute, we consolidate the loading of each expert's weights and biases and maximize their reuse. Furthermore, with ping-pong buffering, we can completely hide the loading latency of most expert MLPs, except in the case of the first expert in the metaqueue or any cases of workload imbalance (where the preceding expert has only a few tokens to compute and finishes its computation early while the next expert is still loading).

\subsection{Unified Linear Layer Module}
\label{sec:unified-linear}

Across the entire \mvit{} model, linear layers with different input/output dimensions are present in the following blocks:
(1) on dense inputs, from input dimension to ViT block hidden dimension; (2) on dense inputs, from ViT block hidden dimension to output dimension; (3) on sparse inputs, from input dimension to MoE block hidden dimension; (4) on sparse inputs, from MoE block hidden dimension to output dimension; (5) on dense inputs, from input dimension to output dimension.
Letting each linear layer consume its own dedicated computing hardware will result in a large resource waste, particularly DSPs, which will largely constrain the maximum parallelism and harm latency. We consolidate all linear layers throughout the model into one single linear layer computation module, which enables all linear layers to take advantage of high parallelism. 
As shown in Fig.~\ref{fig:m3vit}, the unified linear module (green block on the right) can process multiple types of linear layers with flexible run-time configuration, including variable input/output dimension, dense or sparse input, and whether to use GELU.

\noindent
\textbf{Handling variable input/output dimensions}.
Note that it is non-trivial to share resource for linear layers with varying input and output dimensions. In HLS, a linear layer is usually implemented as a nested loop, where the outer loop iterates over the output dimension (\texttt{out\_dim}) and the inner loop iterates over the input dimension (\texttt{in\_dim}). HLS can create one pipeline from both loops only when the inner loop has a constant bound (i.e., \texttt{in\_dim} is constant). But since the linear layers in Edge-MoE have differing input dimensions, \texttt{in\_dim} cannot be constant. Thus the loop cannot be flattened into one pipeline, resulting in severe delays, as an entire output dimension must be processed before the next can start.

\begin{figure}
    \centering
\begin{lstlisting}
int next_i = 0, next_j = 0, iters = out_dim*in_dim;
for (int iter = 0; iter < iters; iter++) {
  int i = next_i, j = next_j;
  next_i = (j == out_dim-1) ? next_i+1 : next_i;
  next_j = (j == in_dim-1) ? 0 : next_j+1;
  // ...loop body...
}
\end{lstlisting}
    \caption{The pseudocode of a manually flattened loop. This technique turns a nested loop (e.g., for-loop \texttt{i} = 0 to \texttt{out\_dim} containing for-loop \texttt{j} = 0 to \texttt{in\_dim}) with variable bounds into a single pipelinable loop.}
    \label{fig:pseudocode-manual-flatten}
\end{figure}

To address this limitation, we propose to use a single manually flattened pipelined loop that tracks virtual loop indices in separate registers, which are incremented manually inside the loop. Figure~\ref{fig:pseudocode-manual-flatten} shows the pseudocode of a manually flattened loop.

The unified linear layer accepts weights and biases in a blocked format, where a single block represents all weights and biases needed for a single cycle of computation. A unified weight loading module is used to read weights from their sequential form from off-chip DRAM into BRAM in the required blocked format.
The unified module consists of a streaming dataflow supporting inter-token pipelining to minimize latency.

\begin{figure}
    \centering
    \includegraphics[width=\linewidth]{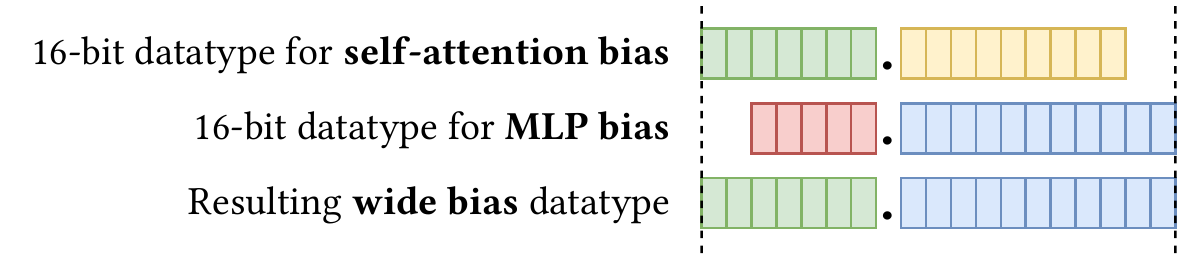}
    \caption{The widened bias fixed-point datatype used in the unified linear layer module.}
    \label{fig:wbias}
\end{figure}

\noindent
\textbf{Handling hybrid fixed-point quantization schemes.}
While unifying all linear layers, the bias loading modules are also unified. However, different layers need to use different fixed-point datatypes for biases to maintain accuracy.
For instance, the linear layers used in self-attention require 16-bit biases with 7 integer bits, but the MLPs in the ViT and MoE blocks require higher precision but lower range, using 16-bit biases with only 5 integer bits.
To utilize the unified linear module, we separately convert these biases to a single wider bias type, which has enough integer bits to cover the range and enough fractional bits to cover the precision of both datatypes, as shown in Figure~\ref{fig:wbias}.

\noindent
\textbf{Handling sparse and dense inputs.}
To be able to handle both sparse and dense inputs simultaneously for the MoE and non-MoE layers, the unified linear layer directly controls the processes reading and writing from DRAM.
The reader and writer processes each contain two submodules for direct (dense) and indirect (sparse, indexed by per-expert queues) DRAM accesses, and the top-level module passes a flag to indicate which should be used at any given time. The indirect writer submodule also supports a weighted accumulation of the linear layer output atop the existing output buffer rather than overwriting it, thus enabling accumulation of per-expert outputs directly without an additional aggregation step.

The DRAM reading and writing processes support aggregation and disaggregation of the blocks processed by the core computation submodule, enabling the use of larger block sizes than supported by the AXI interface to DRAM. 

Finally, a flag controls whether the writer process should apply the GELU function before writing the outputs, which allows the MLPs in the ViT and MoE blocks to incorporate the activation function without an extra step. Thanks to the use of pipelining, the latency added by this activation function is negligible.

\subsection{Gating Network for Multi-Task}

The use of Mixture-of-Experts is particularly important in \mvit{}, as it is the key to enabling efficient Multi-Task Learning. Figure~\ref{fig:m3vit} (left) demonstrates how \mvit{} uses MoE in a multi-task scenario: separate gating networks are used for each task in order to select the best experts for a given combination of token and downstream task.
Our implementation loads gating network weights from DRAM only as needed for computation, which enables easy zero-overhead switching between tasks simply by updating the pointer to the task-specific gating network.

\section{Experiments}

\begin{table}
    \centering
    \caption{Prevailing ViT models and multi-task \mvit{} with mixture-of-expert (MoE) evaluated in this work, and their latency reduction by applying the proposed techniques.}
    \label{tab:model-features}
    \footnotesize
    \setlength{\tabcolsep}{1.1pt}
    \begin{tabular}{c|c|c|c|c|c|c|c|c}
    \toprule
    && \multicolumn{2}{c|}{\textbf{Dimensions}} &&& \multicolumn{2}{c|}{\textbf{Latency (ms)}} & \\
    \textbf{Model} & {\scriptsize \textbf{Layers}} & {\scriptsize \textbf{Hidden}} & {\scriptsize \textbf{MLP}} & {\scriptsize \textbf{Heads}} & {\scriptsize \textbf{Params}} & {\scriptsize \textbf{w/o opt.}} & {\scriptsize \textbf{w/ opt.}} & {\scriptsize \textbf{Speedup}} \\ \hline
    ViT-Base~\cite{dosovitskiy2020image}     &  12 & 768 & 3072 & 12 & 86M & 4061.1 & 414.32 & 9.80\texttimes \\
    ViT-Large     & 24 & 1024 & 4096 & 16 & 307M & 14264 & 1450.6 & 9.83\texttimes \\
    ViT-Huge     & 32 & 1280 & 5120 & 16 & 632M & 29502 & 2997.9 & 9.84\texttimes \\
    DeiT-Small~\cite{pmlr-v139-touvron21a} & 12 & 384 & 1536 & 6 & 22M & 1063.6 & 109.00 & 9.76\texttimes \\
    DeiT-Base & 12 & 768 & 3072 & 12 & 86M & 4061.1 & 414.32 & 9.80\texttimes \\
    \mvit{}+MoE~\cite{liang2022m} & 12 & 192 & 768 & 3 & 7M & 353.44 & 34.64 & 10.20\texttimes \\
    
    \bottomrule
    \end{tabular}
\end{table}

\begin{table}
    \centering
    \caption{CPU and GPU comparisons. {The batch size is 1 image.}}
    \label{tab:cpu-gpu-fpga}
    \small
    \setlength{\tabcolsep}{1pt}
    \begin{tabular}{c|c|c|c|c|c}
        \toprule
        \textbf{Device} & \textbf{Latency} & \textbf{Power} & \textbf{Energy} & \textbf{Frequency} & \textbf{\(\Delta_m\) Acc.} \\\hline
        CPU & 169.72 ms & 14.53 W & 2.466 J (4.90$\times$) & 2500 MHz & +0.76\% \\
        GPU & 13.73 ms & 82.24 W & 1.129 J (2.24$\times$) & 1800 MHz & +0.76\% \\
        \textbf{Edge-MoE} & 34.64 ms & 14.54 W & 0.504 J (1.00$\times$) & 300 MHz & +0.67\% \\
        \bottomrule
    \end{tabular}
\end{table}

\begin{table*}
    \centering
    \caption{Ablation study of our proposed techniques. All the latency, resource, and accuracy values are measured on-board. The baseline is a fully functional \mvit{} accelerator design without our proposed techniques. Applying all six techniques can result in more than 18$\times$ speedup and no accuracy drop.}
    \label{tab:ablation}
    \small
      \setlength{\tabcolsep}{3pt}
    \begin{tabular}{l|c|c|c|c|c|c|c|c}
    \toprule
        && \multicolumn{4}{c|}{\textbf{Hardware resources}} & \multirow{2}{*}{\textbf{\begin{tabular}{@{}c@{}}Sem.\ seg.\\(mIoU \(\uparrow\))\end{tabular}}} & \multirow{2}{*}{\textbf{\begin{tabular}{@{}c@{}}Depth\ est.\\(RMSE \(\downarrow\))\end{tabular}}} & \multirow{2}{*}{\textbf{{\begin{tabular}{@{}c@{}}MTL acc.\\gain \(\Delta_m\) (\(\uparrow\))\end{tabular}}}} \\
        \textbf{Architecture} & \textbf{Latency (Speedup)} & \textbf{BRAM} & \textbf{DSP} & \textbf{LUT} & \textbf{FF} &&& \\
        \hline
        Baseline w/o our proposed techniques & 650.3 ms (1.00\texttimes{}) & 84.6\% & 65.9\% & 60.8\% & 46.9\% & {63.8066} & {0.0373} & {+0.58\%} \\
        + Expert-by-expert reordering (\S\ref{sec:expert-reorder}) & 433.4 ms (1.50\texttimes{}) & 84.3\% & 65.9\% & 59.0\% & 45.4\% & {63.8066} & {0.0373} & {+0.58\%} \\
        + Single-pass dynamic bias softmax (\S\ref{sec:softmax}) & 353.4 ms (1.84\texttimes{}) & 84.0\% & 67.9\% & 59.0\% & 45.7\% & {63.8066} & {0.0373} & {+0.58\%} \\
        + Accurate, low-cost GELU (\S\ref{sec:gelu}) & 212.9 ms (3.05\texttimes{}) & 64.1\% & 63.0\% & 45.5\% & 34.9\% & {63.8491} & {0.0372} & {+0.67\%} \\
        + Unified linear layer module (\S\ref{sec:unified-linear}) & 104.3 ms (6.23\texttimes{}) & 47.5\% & 57.3\% & 44.1\% & 27.0\% & {63.8491} & {0.0372} & {+0.67\%} \\
        + Attention reordering on \(Q \times K\) (\S\ref{sec:attention-reorder}) & 59.2 ms (10.98\texttimes{}) & 48.1\% & 61.6\% & 46.5\% & 27.5\% & {63.8491} & {0.0372} & {+0.67\%} \\
        + Attention reordering on \(M' \times V\) (\S\ref{sec:attention-reorder}) & 34.6 ms (18.77\texttimes{}) & 50.1\% & 76.3\% & 46.8\% & 29.4\% & {63.8491} & {0.0372} & {+0.67\%} \\
        \midrule
        \multicolumn{6}{r|}{Software baseline accuracy (PyTorch)} & {63.9450} & {0.0372} & {+0.76\%} \\
    \bottomrule
    \end{tabular}
\end{table*}

We conduct several \textit{on-board, verified} experiments to demonstrate the effectiveness of our proposed methods.
All on-board code is implemented through High-Level Synthesis through Xilinx Vitis HLS 2021.1. We deploy bitstreams to Xilinx ZCU102 FPGA and use the PYNQ library for host code. The clock frequency of our design is 300 MHz. All experiments use the Cityscapes dataset~\cite{Cordts2016Cityscapes}, with images of size 128$\times$256 split into patches of size 16$\times$16.

Our main comparison is with CPU and GPU baselines and the FPGA design without our proposed techniques. We find it difficult to make a fair comparison with prior works for two main reasons. First, we target ViT acceleration from a different angle than prior works and do not rely on model compression or re-training; these techniques are orthogonal to ours and can be applied together. Second, prior works in ViT acceleration lack an end-to-end on-board implementation, which exposes difficulties to compare with.

\subsection{Comparison with CPU and GPU on \mvit{}}

We first evaluate our accelerator for \mvit{} since it is the state-of-the-art multi-task ViT model,
which uses standard ViT as its backbone but also is equipped with MoE features, making it more challenging than standard ViT models. Therefore, we choose \mvit{} as our primary baseline, evaluated on the semantic segmentation and depth estimation tasks.
We deploy our accelerator onto ZCU102 FPGA and compare against CPU (Intel Xeon 6226R) and GPU (NVIDIA RTX A6000) baselines implemented in PyTorch using FastMoE~\cite{fastmoe}, a library that uses scatter/gather techniques for optimized MoE computation. Power is measured using Intel RAPL and NVIDIA SMI, respectively.
We use 16-bit fixed-point weights and 32-bit fixed-point activations. Results are in Table~\ref{tab:cpu-gpu-fpga}.

First, in terms of accuracy, both the software baseline \mvit{} and Edge-MoE outperform single-task learning (STL) baselines (\(\Delta_m\) Acc. in Table~\ref{tab:cpu-gpu-fpga}): software \mvit{} outperforms STL by +0.76\%, and Edge-MoE on FPGA performs +0.67\% better -- a drop of only 0.09\%. 
Second, in terms of energy efficiency, Edge-MoE on FPGA achieves significant savings: 4.90$\times$ over CPU and 2.24$\times$ over GPU.

\subsection{Standard ViT Models}

We also evaluate how well our optimization techniques can reduce the latency of several different ViT models when implemented on FPGA.

Results in Table~\ref{tab:model-features} show consistent improvement from our proposed methods ranging from 9.76$\times$ to 9.84$\times$ on non-\mvit{} models, as well as over 10$\times$ speedup on \mvit{} with Mixture-of-Experts. The reliable speedup gained from our optimizations shows that, besides our model-specific techniques for \mvit{}, our methods are generally applicable and effective when applied to any ViT model.

\subsection{Latency and Resource Breakdown}

\begin{figure}
    \centering
    \includegraphics[width=0.49\linewidth]{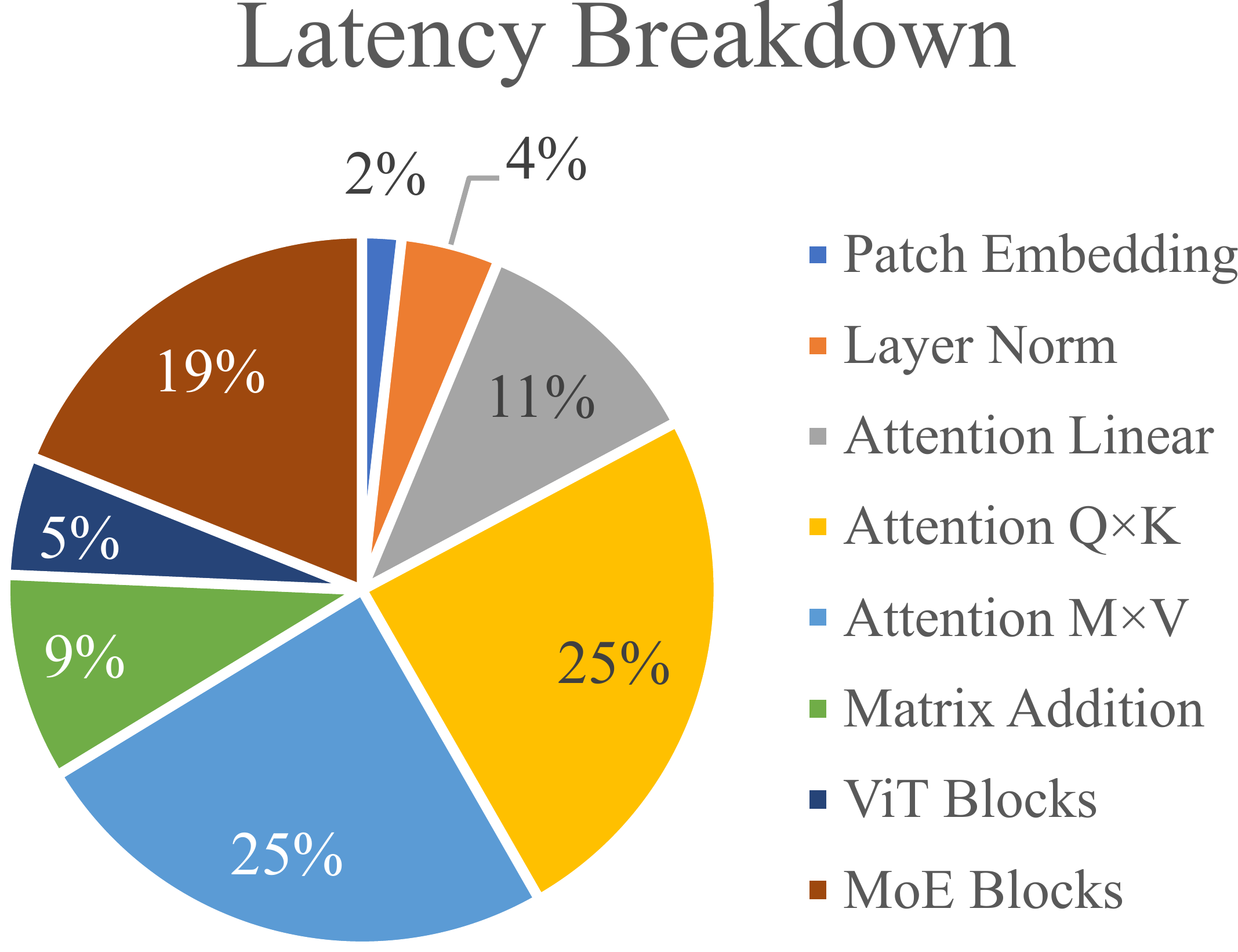}\hspace{2pt}
    \includegraphics[width=0.49\linewidth]{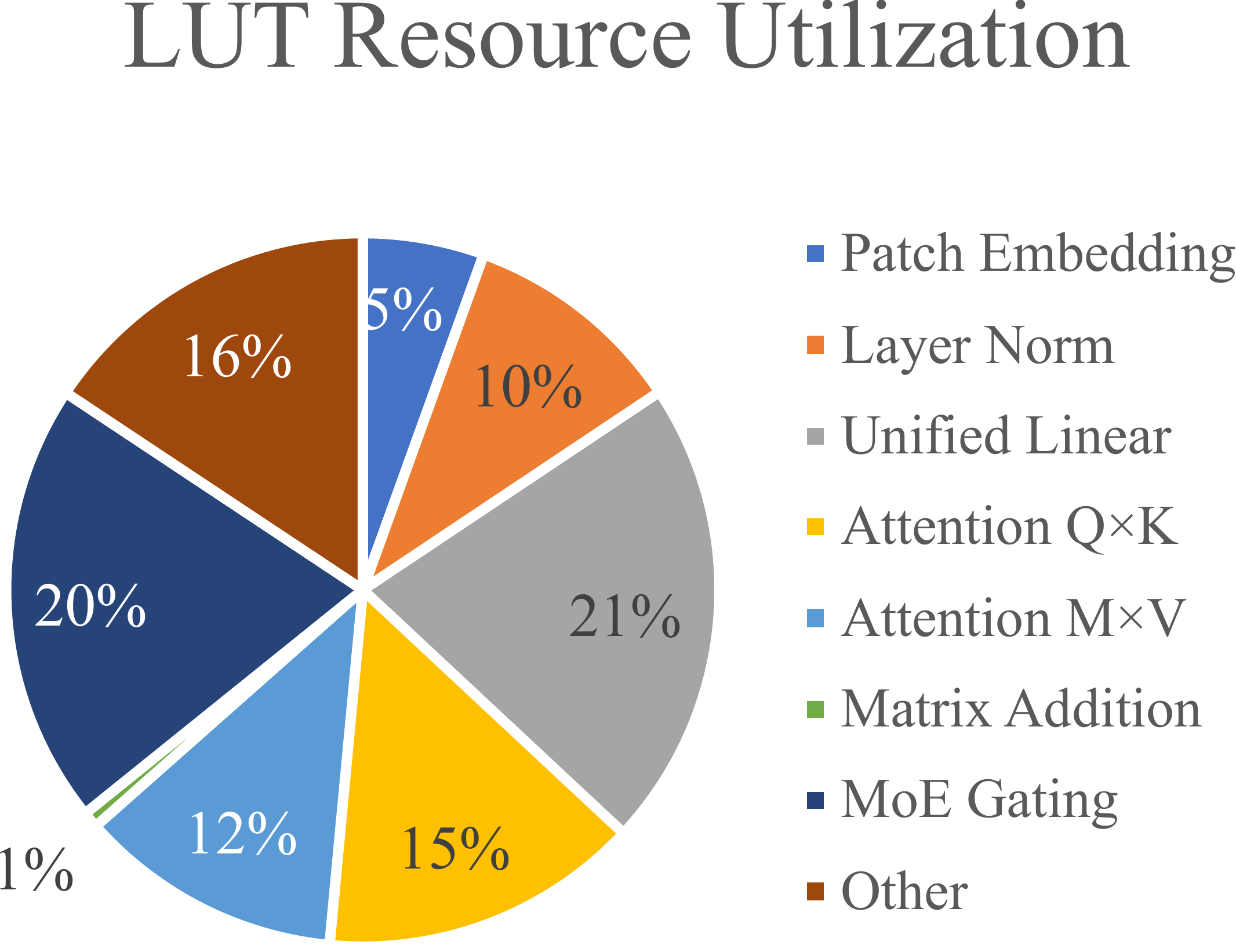}
    \caption{A breakdown of latency and LUT usage in our FPGA implementation of the \mvit{} model.}
    \label{fig:latency-breakdown}
\end{figure}

To understand which parts of our model are the most expensive and which take the most time to compute, we measure the on-board latency and resource usage of the different components of our \mvit{} implementation. Figure~\ref{fig:latency-breakdown} displays our findings.

Even at 4\texttimes{} parallelism, the attention multiplications \(Q \times K\) and \(M' \times V\) take half of the total computation time, demonstrating the necessity of accelerating this computation.

Additionally, the effectiveness of our unified linear layer stands out: it takes the largest portion of LUT resources, but as a result, it greatly accelerates the attention linear layers, ViT blocks, and MoE blocks, which take only 35\% of the overall latency \textit{combined.}

\subsection{Ablation Study}

In Table~\ref{tab:ablation}, we conduct an ablation study to evaluate the effect of our optimizations on latency, resource usage, and accuracy. We show the incremental progression of these factors as each of our key techniques are applied, including the indirect effect of lower-cost hardware resources enabling greater increases in parallelism.

We find that each of our features leads to a noticeable speedup. Additionally, most hardware resources follow a downward trend over time, demonstrating empirically that our techniques are hardware-friendly. However, the number of DSPs used remains roughly the same throughout most of the ablation study, as consistent increases in parallelism roughly negated the effect on the number of DSPs.

We call attention to three rows of Table~\ref{tab:ablation} in particular:

\subsubsection{Single-pass softmax with dynamic bias}
Due to the challenges caused by overflow of \(\exp(\cdot)\) described in Sec.~\ref{sec:challenge-softmax}, all architectures use a dynamic bias, as in Sec.~\ref{sec:dynamic-bias}. However, the third row introduces our single-pass softmax optimization from Sec.~\ref{sec:single-pass-softmax}, rather than three passes.

\subsubsection{Accurate, low-cost GELU}
Most of our optimizations do not affect the downstream task accuracy at all and are mathematically equivalent to their unoptimized versions. However, our accurate, low-cost GELU implementation supersedes a less accurate sigmoid-based approximation from previous architectures, which is described in Sec.~\ref{sec:challenge-gelu}. Not only does our calibration-based approach increase the approximation accuracy, it also further reduces resource usage, allowing for higher parallelism which results in 1.66$\times$ speedup over the previous architecture.

\subsubsection{Unified linear layer module}
Our proposed unified linear layer module led to the single largest relative change in latency in the table, over 2$\times$ faster than the previous architecture, while also reducing the usage of all four hardware resources. As described in Sec.~\ref{sec:unified-linear}, the unified linear layer supersedes five dedicated hardware modules for separate linear layers, thereby dramatically reducing the number of resources consumed. This, in turn, allowed for significantly higher parallelism in the unified linear layer, which benefited all parts of \mvit{} that were previously using a dedicated module, while still leaving reduced resource utilization.

\section{Conclusion}

In this paper, we proposed a novel FPGA accelerator with a collection of innovative techniques for Vision Transformer (ViT) models and a state-of-the-art multi-task ViT using Mixture-of-Expert (MoE). To the best of our knowledge, this is the first end-to-end FPGA implementation for multi-task ViT, with verified functionality on-board. We propose five key techniques: (1) an attention computation reordering mechanism which reduces the bandwidth requirement from proportion to constant, regardless of parallelism; (2) a single-pass softmax approximation achieving both high accuracy and speed; (3) an accurate GELU activation approximation with extremely low hardware cost; (4) a unified and flexible linear layer for aggressive resource sharing; (5) a novel expert-by-expert MoE computing mechanism requiring zero memory and data transfer overhead. The proposed techniques can reduce latency by more than $18\times$ applying to \mvit{} and more than $9\times$ applying to standard ViT models. Comparing with CPU and GPU, our design achieves $4.90\times$ and $2.24\times$ better energy efficiency, respectively. On the multi-task self-driving car dataset, we achieve nearly 30 frames per second, showing promising performance for MTL ViT in real-world applications.

\section*{Acknowledgements}

This work and its authors are partially supported by the 2022 Qualcomm Innovation Fellowship program, the National Science Foundation under Grant No.\ 2202329, and the Center for Research into Novel Computing Hierarchies (CRNCH) at Georgia Tech.

\bibliographystyle{IEEEtranS}
\bibliography{refs}

\end{document}